\documentclass[draftclsnofoot,onecolumn]{IEEEtran}
\IEEEoverridecommandlockouts
\usepackage{amsmath,graphicx,amssymb,mathtools,bm}
\usepackage{subfigure}
\usepackage{hyperref}
\usepackage{cite}
\usepackage{amsmath,amssymb,amsfonts}
\usepackage{algorithmic}
\usepackage{textcomp}
\usepackage{xcolor}
\usepackage[justification=centering]{caption}
\usepackage{verbatim}  
\usepackage{bm}  
\usepackage{mathrsfs} 
\usepackage{algorithm} 
\usepackage{algorithmic} 
\usepackage{booktabs}
\usepackage{textcomp}  
\usepackage{multirow}  
\usepackage{lettrine}   
\usepackage{color}  
\usepackage{amsmath}
\usepackage{amssymb}

\def\BibTeX{{\rm B\kern-.05em{\sc i\kern-.025em b}\kern-.08em
    T\kern-.1667em\lower.7ex\hbox{E}\kern-.125emX}}
\begin{document}

\title{Bayesian Optimization-Based Beam Alignment for MmWave MIMO Communication Systems
}

\author{Songjie Yang,
	Baojuan Liu,
	Zhiqin Hong,
	Zhongpei Zhang,~\IEEEmembership{Member,~IEEE},
	
	\thanks{This work was supported in part by the National Key Research and Development Program of China under Grant 2020YFB1806805. (\textit{Corresponding author:
			Zhongpei~Zhang}.)
	}
	
	\thanks{Songjie Yang, Baojuan Liu, Zhiqin Hong and Zhongpei Zhang are with the National Key Laboratory of Science and Technology on Communications, University of Electronic Science and Technology of China, Chengdu 611731, China. (e-mail:
		yangsongjie@std.uestc.edu.cn; baojuanl@yeah.net ;202021220520@std.uestc.edu.cn; zhangzp@uestc.edu.cn).

	}
}

\maketitle

\begin{abstract}
Due to the very narrow beam used in millimeter wave communication  (mmWave), beam alignment (BA) is a critical issue. In this work, we investigate the issue of mmWave BA and present a novel beam alignment scheme on the basis of a machine learning strategy, Bayesian optimization (BO). In this context, we consider the beam alignment issue to be a black box function and then use BO to find the possible optimal beam pair. During the BA procedure, this strategy exploits information from the measured beam pairs to predict the best beam pair. In addition, we suggest a novel BO algorithm based on the gradient boosting regression tree model. The simulation results demonstrate the spectral efficiency performance of our proposed schemes for BA using three different surrogate models. They also demonstrate that the proposed schemes can achieve spectral efficiency with a small overhead when compared to the orthogonal match pursuit (OMP) algorithm and the Thompson sampling-based multi-armed bandit (TS-MAB) method.
\end{abstract}

\section{Introduction}
 There has been a surge of interest in Millimeter wave (mmWave) communications which are promising to provide high data rate. However, the propagation loss at mmWave frequencies is substantially larger. It is well established that directional communications come in handy for overcoming the severe attenuation. For all that, it is noteworthy that narrow beams in directional communications are highly susceptible to mis-alignment, hence they require huge overhead for the beam alignment (BA) process. There is evidence that BA plays a crucial role in establishing the mmWave communication link. Thus, it is of paramount significance to develope efficient BA strategies.

In general, BA is used to determine the optimal pair of analog precoder and combiner which attains the maximum received signal strength (RSS). On the one hand, previous research on BA can be broadly classified into two categories: (i) Sequential Search and (ii) Hierarchical Search. \cite{SIDE1} and \cite{SIDE2} proposed a side search scheme which made use of quasi-omnidirectional beam pattern at the transmitter (TX), while the receiver (RX) made an exhaustive scan over the beam space; then this procedure was reversed in the second stage with the TX scanning the beam space, while the RX adopted a quasi-omnidirectional beam to receive. However, adopting an omnidirectional beam for BA requires a huge power for a large coverage and induces interference to others. Another attractive approach is the use of generating multiple directional beams simultaneously. For instance, the authors in \cite{beamcoding} generated multiple beams simultaneously by manipulating the antenna weights and then coded the beams via a unique signature. Furthermore, the agile-link beam training scheme in \cite{alige} adopted randomized hashing and sparse fast Fourier transform (FFT) to generate multiple beams and then used a voting method to find the optimal direction. For all that, transmitting multiple beams simultaneously is difficult and it suffers from high power in the side lobes. In addition, the hierarchical search strategies apply a combination of high and low-resolution antenna patterns iteratively for beam training, i.e., the TX first makes an exhaustive search on wide beams and then conducts a more precise search on narrow beams. In this regard, it is of paramount significance in designing hierarchical codebooks \cite{H1,H2}.

 On the other hand, some researches documented other approaches to beam alignment such as compressing sensing (CS), side information (SI) and machine learning (ML). CS theory captures sparse paths in a normalized angular domain to convert the sparse path estimation problem to a virtual angle-of-arrival (AoA) and angle-of-departure (AoD) estimation \cite{CS1,CS2}. Additionally, SI based methods exploit position information \cite{SI1}, orientation information \cite{SI2}, auxiliary array information \cite{SI3} or spatial information extracted from sub-6 GHz channels \cite{SI4} to reduce the BA overhead or improve the performance.
 
 Recently, a considerable literature has grown up around the theme of ML, which can answer the issue of whether it is feasible to get useful information to minimize the beam space for the subsequent beam alignment procedure given the measurements of previous beam pairings. 
 Several studies \cite{MAB2,MAB3,MAB4,MAB5} investigated contextual multi-armed bandit (MAB) for BA, where each beam pair index is regarded as an arm. In MAB problems, there is a decision maker shaking a subset of arms of unknown expected rewards with the target to maximize the cumulative reward over time (one arm may be repeatedly shaken).
 In \cite{MAB2}, the position-aided MAB corrected the positioning error to make the BA performance better. \cite{MAB3} proposed a general contextual MAB for the TX beam alignment. In this regard, the authors regarded the AoA of the RX as the context information supplied to the TX for the BA learning algorithm. Chafaa et al. \cite{MAB4} framed the BA problem as a distributed structure between the TX and the RX and adopted adversarial MABs without assuming an underlying channel distribution. \cite{MAB5} documented a correlation structure MAB which exploited the correlation information derived from nearby beams for choosing the next beam efficiently to accelerate BA.
 
However, since we are primarily concerned with the end-game reward rather than the cumulative reward, this will result in unnecessary overhead for MAB-based BA. In addition, MAB is inapplicable for non-contextual BA, i.e., there is no position or other information. Furthermore, since MAB-based and beam sweeping-based schemes depend on the size of codebook, they ignore that beam alignment is a continuous optimization issue. 
Motivated by the above reasons, we present a more appropriate machine learning method for BA.

In this study, a novel BA scheme adopting Bayesian optimization (BO) algorithm is proposed right after fomulating BA as a black box problem. To our best knowledge, this is the first paper devoted to resolving BA with BO. Indeed, BO is a powerful technique for exploring the extrema of the objective function which is expensive or hard to evaluate. It works flawlessly when these evaluations are prohibitively costly, when the problem at hand is a black box function, or when one doesn't have effective access to derivatives \cite{BOS1,BOS2}. 
It is general in global optimization, and true for the BA problem, that the objective function is a black box and expensive to evaluate: there is no an mathematical expression for us to analyze the optimal beam pair without knowing the channel information, but only spending time slots sweeping beams to find the optimal beam pair.
It is a general property of global optimization, and one that holds true for our BA problem, that the objective function is a black box and expensive to evaluate: there is no mathematical expression that allows us to analyze the optimal beam pair without knowing the channel information, but only by sweeping beam codebooks. 

 In this sense, we apply BO to facilitate beam alignment in two regards: 1) a $\emph{surrogate model}$ for estimating the objective function with AoA, AoD and RSS based on the existing samples, and 2) an $\emph{acquisition function}$ for selecting the next sample point (beam pair) at which to evaluate the objective function. For concreteness, we model the BA procedure as a black box function with AoA and AoD input and RSS output, and then adopt the common BO strategy based on Gaussian Process (GP) and Sequential Model-based Algorithm Configuration (SMAC) surrogate models for BA respectively. Following that, we propose a BO strategy with the gradient boosting regression tree (GBRT) model to cope with the BA problem. 

\emph{Notations:} We use the following notations throughout this paper. $\vert\cdot\vert$ represents the modulus function, $\Vert\cdot\Vert$ denotes the $\ell_2$ norm, $(\cdot)^{\rm T}$, $(\cdot)^{\rm H}$ and $(\cdot)^{-1}$ denote transpose, conjugate transpose and inversion respectively. $\mathcal{N}(\mathbf{a},\mathbf{A})$ or $\mathcal{CN}(\mathbf{a},\mathbf{A})$ denotes a real or complex Gaussian vector with mean $\mathbf{a}$ and covariance matrix $\mathbf{A}$. $\mathbb{C}^{x\times y}$ represents the complex-value matrices with the space of $x\times y$. $\mathbf{I}_N$ is the $N\times N$ identity matrix. Finally, expectation is denoted by $\mathbb{E}[\cdot]$.
\section{System And Channel Model}

	\begin{figure}
	\centering
	\includegraphics[width=8.45cm,height=4.86cm]{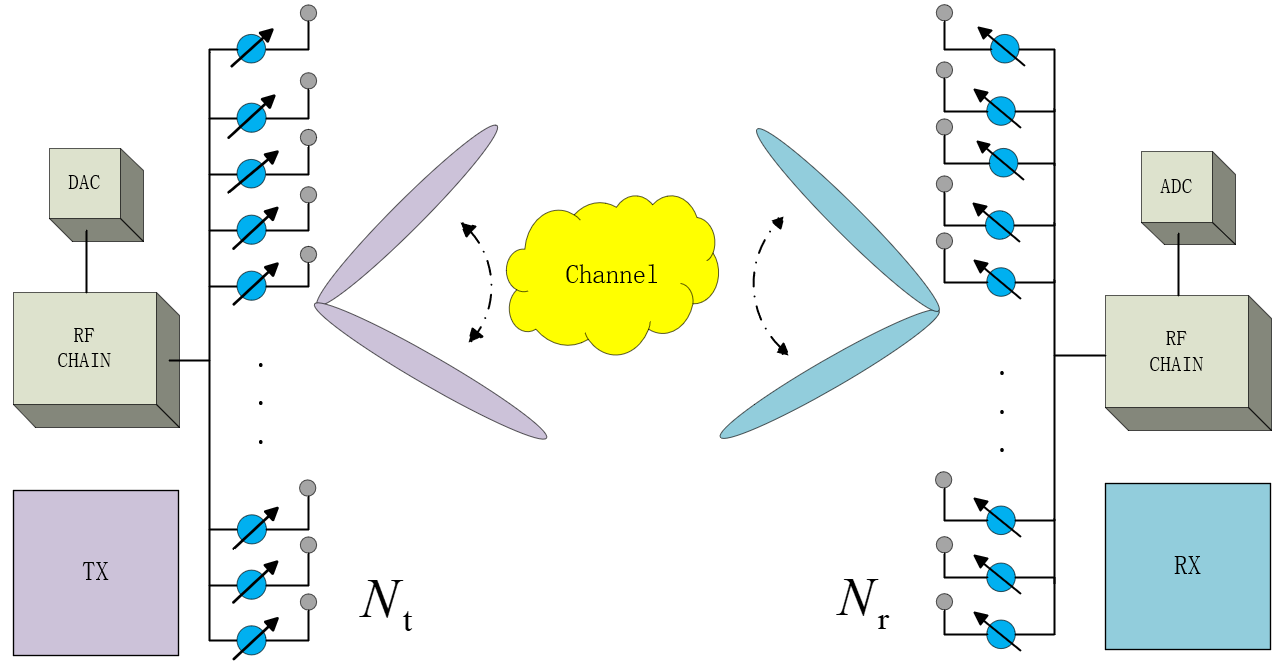}
	
	\caption{Structure of the mmWave system with one RF chain, where the TX (RX) are equipped with $N_t$ ($N_r$) antennas.}
\end{figure}
\figurename{1} exhibits a point-to-point mmWave communication system, where the TX (RX) is equipped with $N_t (N_r)$ antennas. Postulate that the length of the BA phase is $K$ and  the received signal at the RX at the time $k$ is $y[k]$, then we have
\begin{equation}\label{x}
	y[k]=\sqrt{P_t}\mathbf{u}[k]^{\rm H}\mathbf{H}\mathbf{v}[k]s[k]+\mathbf{u}[k]^{\rm H}\mathbf{w}[k], \ k<K,
\end{equation}
where $P_t$ is the average transmit power of the TX, $s[k]$ is the pilot signal such that $\vert s[k] \vert^2=1$, $\mathbf{v}[k]\in\mathbb{C}^{N_t\times1}$ is the TX beamforming vector with $\vert \mathbf{v}[k] \vert^2=1$, $\mathbf{u}[k]\in\mathbb{C}^{N_r\times 1}$ is the RX beamforming vector with $\vert \mathbf{u}[k] \vert^2=1$, $\mathbf{H}\in\mathbb{C}^{N_r\times N_t}$ is the channel matrix and $\mathbf{w}[k]$ represents the independent and identically distributed (i.i.d) additive white Gaussian noise following $\mathcal{CN}(0,\sigma_n^2\mathbf{I}_{N_r})$.

As we all know, due to the limited number of scatterers in the mmWave propagation environment, the mmWave channel cannot follow the rich scattering model of low-frequency assumption. So far, the geometric Saleh-Valenzuela channel model has been widely adopted for characterizing the spatial correlation of mmWave channels \cite{SV}. In general, the channel matrix can be represented as
\begin{equation}
	\mathbf{H}=\sqrt{\frac{N_tN_r}{L}} \sum_{l=1}^L\alpha_l\mathbf{a}_r(\varphi_l)\mathbf{a}^{\rm H}_t(\theta_l),
\end{equation}
where $\alpha_l$ denotes the complex gain of the $l$-th path, $\theta_l$ and $\varphi_l$ are AoD and AoA of the $l$-th path respectively. $\mathbf{a}_t(\cdot)$ and $\mathbf{a}_r(\cdot)$ are the steering vectors at the TX and the RX for uniform linear arrays (ULA), respectively. The array response of the TX and the RX can be written by
\begin{equation}
	\mathbf{a}_t(\theta)=\frac{1}{N_t}[ 1,e^{j2\pi dsin\theta /\lambda},\cdots,e^{j2\pi(N_t-1) dsin\theta /\lambda}]^{\rm T},
\end{equation}
\begin{equation}
	\mathbf{a}_r(\phi)=\frac{1}{N_r}[ 1,e^{j2\pi dsin\phi /\lambda},\cdots,e^{j2\pi(N_r-1) dsin\phi /\lambda}]^{\rm T},
\end{equation}
where $d$ is the antenna inter-element spacing, $\lambda$ is the antenna wavelength.
\section{Problem Formulation And Proposed Scheme }
It is well known that the BA problem is to design $\mathbf{v}$ and $\mathbf{u}$ to maximize $\vert\mathbf{u}^{\rm H}\mathbf{H}\mathbf{v}+w\vert^2$ in (\ref{x}). However, this is a black box function  given the channel matrix is unknown. Generally, this problem is addressed by first establishing the beam codebook and then sweeping beams. Consider over-complete dictionaries $\mathbf{A}_{\rm R}\in\mathbb{C}^{N_r\times G_r}$ and $\mathbf{A}_{\rm T}\in\mathbb{C}^{N_t\times G_t}$, we have
\begin{equation}\label{D_A}
	\mathbf{A}_{\rm R}=[{\mathbf{a}}_{r}(-1),{\mathbf{a}}_{r}(-1+\frac{2}{G_r}),\cdots,{\mathbf{a}}_{r}(1-\frac{2}{G_r})],
\end{equation}
\begin{equation}
	\mathbf{A}_{\rm T}=[{\mathbf{a}}_{t}(-1),{\mathbf{a}}_{t}(-1+\frac{2}{G_t}),\cdots,{\mathbf{a}}_{t}(1-\frac{2}{G_t})],
\end{equation}
where $G_r$ and $G_t$ are the number of the AoAs and AoDs on the grid. 
With the increment of the size of codebooks, the BA procedure will achieve a considerable performance. However, this will incur an unacceptable training overhead for mmWave communications.

In this section, we adopt BO based on different surrogate models to maximize the black box function to attain the optimal AoA and AoD.
The core idea of BO is to specify a prior belief for the possible objective function, and then sequentially refine the model as sample data are observed by Bayesian posterior updating. Furthermore, the Bayesian posterior characterizes our updated beliefs---given sample space---on the likely objective function we are optimizing.

In particular, we consider the BA problem where the optimal value is sought for an expensive function $f: \mathcal{Z}\in[-\pi/2,\pi/2]$,
\begin{equation}\label{BA}
	\begin{aligned}
	(\theta^{\star},\varphi^{\star})=& \underset{\theta,\varphi\in\mathcal{Z}}{\rm arg \ max} \ f(\theta,\varphi) \\
	=& \underset{\theta,\varphi\in\mathcal{Z}}{\rm arg \ max} \ \vert\mathbf{a}^{\rm H}_r(\varphi)\mathbf{H}\mathbf{a}_t(\theta)+w\vert^2.
	\end{aligned}
\end{equation}

Fundamentally, BO is a sequential model-based approach to coping with problem (\ref{BA}). Algorithm \ref{BO} exhibits the procedure of BO which starts by initializing a surrogate model $\mathcal{M}$ with a small scale samples from the domain $\mathcal{Z}$.
Denoting the beam pair $(\theta,\varphi)$ by $\mathbf{z}$. 
 As we accumulate observations $\mathcal{D}_m=\{ (\mathbf{z}_1,f(\mathbf{z}_1)),(\mathbf{z}_2,f(\mathbf{z}_2)),\cdots,(\mathbf{z}_m,f(\mathbf{z}_m))\}$ through randomly selecting $m$ beam pairs from the codebook by the transceiver, a prior distribution $p(f)$ is combined
with the likelihood function $p(\mathcal{D}_m|f)$ to induce the posterior distribution
\begin{equation}
	p(f|\mathcal{D}_m) \propto p(\mathcal{D}_m|f)p(f),
\end{equation}
where the posterior $p(f|\mathcal{D}_m)$ captures the updated beliefs about the unknown objective function $f$.

Although the initial samples are used to preliminarily fit the objective function, the next sampling point (beam pair) is supposed to be selected to update the posterior probability to make the performance better, and how to select the next optimal sampling point is implemented by the acquisition function $\mathcal{L}$ which leverages the uncertainty in the posterior to make exploration, that is, we can exploit $\mathcal{L}$ to attain the new next observation point and feedback to the TX\footnote{In this work, error- and delay-free feedback with very few bits is assumed \cite{MAB6}. Indeed, a feedback-reduced BO scheme is that observe multiple objective function values using $\mathcal{L}$ rather than one at a time.}. Following that, $p(f|\mathcal{D}_{m+1})$ is updated according to new data space. Furthermore, the BO process iterates $n$ times in this manner to achieve high performance.
\begin{algorithm}[!t] 
	\caption{Bayesian Optimization-Based Beam Alignment} 
	\label{BO}      
	\begin{algorithmic} [1]
		\footnotesize{
		
			\STATE{$\textbf{Initialize:}$ Randomly select $m$ beam pairs to construct the sample space $\mathcal{D}_m=\{ (\mathbf{z}_1,f(\mathbf{z}_1)),(\mathbf{z}_2,f(\mathbf{z}_2)),\cdots,(\mathbf{z}_m,f(\mathbf{z}_m))\}$
			}

			\FOR{$i=1,2,\cdots,n$
			}
		
			\STATE { 
				Select new beam pair  $\mathbf{z}_{m+i}$ by optimizing acquisition function $\mathcal{L}$: \\ $\mathbf{z}_{m+i}=\underset{\mathbf{z}\in\mathcal{Z}}{\rm arg \ max} \ \mathcal{L}(\mathbf{z};\mathcal{D}_{m+i})$. 
				
			}
				
			\STATE { 
			Query objective function to acquire new observation  $f(\mathbf{z}_{m+i})$ and feedback.
				
			}
			\STATE { 
			Augment sample space $\mathcal{D}_{m+i}=\{ \mathcal{D}_m, (\mathbf{z}_{m+i},f(\mathbf{z}_{m+i}))\}$.
			
		}
		\STATE { 
	Update $p(f|\mathcal{D}_{m+i})$.
		
	}
			\ENDFOR

		}
	\end{algorithmic}
\end{algorithm}

In brief, the BO framework has
two key ingredients: \romannumeral1) a probabilistic surrogate model, consisting of a prior distribution which can capture our beliefs about the behavior of the unknown objective function and an observation model which describes the data generation mechanism, and \romannumeral2) an acquisition function which can represent how optimal a sequence of queries are.

\subsection{Surrogate Model}
There is a variety  of surrogate models for the BO framework, e.g., Gaussian Process (GP) \cite{SM1}, Sequential Model-based Algorithm Configuration (SMAC) \cite{SM3}, Tree-structured Parzen Estimator (TPE) \cite{SM2}. 

For ease of understanding, we first discuss how the common GP surrogate model $\mathcal{M}$ is placed on $f$. After that, we propose a BO scheme with gradient boosting regression tree (GBRT) based surrogate model which is similar to SMAC.
\subsubsection{Gaussian Process}
 GP is an extension of the multivariate Gaussian distribution to the infinite dimensional random process for which any finite combination of dimensions will  be a Gaussian distribution. It has become a common  surrogate model for fitting the objective function in BO. In this regard, the function $f$ is postulated as a implementation of a GP with covariance kernel $\mathcal{K}$ and mean $\bm{\mu}$, i.e.,
\begin{equation}
	f(\mathbf{z})\sim \mathcal{N}(\bm{\mu},\mathcal{K}).
\end{equation}

It is noteworthy that the selection of the kernel function $\mathcal{K}$ in particular can make an effect on the quality of the surrogate reconstruction. For convenience, the common squared exponential function is selected:
\begin{equation}
	\mathcal{K}(\mathbf{z}_i,\mathbf{z}_j)={\rm exp}(-\frac{1}{2}\Vert \mathbf{z}_i-\mathbf{z}_j\Vert^2).
\end{equation}

Given the sample space $\mathcal{D}_m$, the function values are drawn  in the light of a multivariate normal distribution $\mathcal{N}(0,\mathbf{K})$, where the kernel matrix $\mathbf{K}$ is denoted by
\begin{equation}
	\mathbf{K}=
	\begin{bmatrix}
		\mathcal{K}(\mathbf{z}_1, \mathbf{z}_1) & \cdots & 	\mathcal{K}(\mathbf{z}_1, \mathbf{z}_m) \\
		\vdots & \ddots & \vdots
		
		 \\ 
				\mathcal{K}(\mathbf{z}_m, \mathbf{z}_1) & \cdots & 	\mathcal{K}(\mathbf{z}_m, \mathbf{z}_m)
	\end{bmatrix} .
\end{equation}

In our BA optimization problem, we make use of data from a small set of samples during the BA process to fit the GP and attain the posterior. According to the properties of GP, $f(\mathbf{z}_{1:m})$ and $f(\mathbf{z}_{m+1})$ are jointly Gaussian:
\begin{equation}
	\begin{bmatrix}
		f(\mathbf{z}_{1:m}) \\
		f(\mathbf{z}_{m+1}) 
	\end{bmatrix}
	\sim \mathcal{N} 
	\begin{pmatrix}
	0,
	\begin{bmatrix}
		\mathbf{K} & \mathbf{k} \\
		\mathbf{k}^{\rm T} & \mathcal{K}(\mathbf{z}_{m+1}, \mathbf{z}_{m+1})
	\end{bmatrix}
	\end{pmatrix},
\end{equation}
where $\mathbf{k}=[\mathcal{K}(\mathbf{z}_{m+1}, \mathbf{z}_1), \mathcal{K}(\mathbf{z}_{m+1}, \mathbf{z}_2),\cdots, \mathcal{K}(\mathbf{z}_{m+1}, \mathbf{z}_{m+1})]$.

Following that, an expression for the predictive distribution can be derived as
\begin{equation}
	p(f(\mathbf{z}_{m+1})|\mathcal{D}_{m},\mathbf{z}_{m+1})=\mathcal{N}(\bm{\mu}_{m}(\mathbf{z}_{m+1}),\sigma_m^2(\mathbf{z}_{m+1})),
\end{equation}
where $\bm{\mu}_{m}(\mathbf{z}_{m+1})=\mathbf{k}^{\rm T}\mathbf{K}^{-1}f_{1:m}$ and  $\sigma_m^2(\mathbf{z}_{m+1})=\mathcal{K}(\mathbf{z}_{m+1},\mathbf{z}_{m+1})-\mathbf{k}^{\rm T}\mathbf{K}^{-1}\mathbf{k}$.

In other words, $\bm{\mu}_{m}(\cdot)$ and $\sigma_m^2(\cdot)$ describe the sufficient statistics of the posterior distribution function $p(f(\mathbf{z}_{m+1})|\mathcal{D}_{m},\mathbf{z}_{m+1})$.
\subsubsection{Gradient Boosting Regression Tree}
Compared to GP, another choice for the probabilistic regression model is an ensemble of regression trees, like SMAC.  In contrast to the random forest (RF) model applied in SMAC as the surrogate model, we adopt the GBRT model instead. Both RF and GBRT belong to ensemble learning, but their essential difference lies in ensemble strategies. RF is built adopting a strategy called bagging in which each  tree is used as a parallel estimator. However, GBRT employs boosting strategy, which works in a similar way, except that the trees grow in order: each tree grows using the information derived from the previously grown trees.

A common approach in constructing the predictive distribution is to assume that it follows a Gaussian  $\mathcal{N}(f|\hat{\mu},\hat{\sigma}^2)$. The parameters $\hat{\mu}$ and $\hat{\sigma}$ may be chosen as
the empirical mean and variance of the regression values $s$, from the set of regression trees $\mathcal{B}$ in the GBRT model \cite{GBDT}:
\begin{equation}
	\hat{\mu}=\frac{1}{\vert\mathcal{B}\vert}\sum_{s\in\mathcal{B}} \ s(\mathbf{z}),
\end{equation}
\begin{equation}
	\hat{\sigma}^2=\frac{1}{\vert\mathcal{B}\vert-1}\sum_{s\in\mathcal{B}} \ (s(\mathbf{z})-\hat{\mu})^2,
\end{equation}
where $\vert\mathcal{B}\vert$ is the cardinality of $\mathcal{B}$.
\subsection{Acquisition Function}
Thus far, we have discussed placing the priors over the surrogate model and how to update them in line with new observations. Then we will introduce how to select a optimal sample point to obtain the observation by maximizing the acquisition function $\mathcal{L}$. That is, we wish to sample $f$ at $\underset{\mathbf{z}\in\mathcal{Z}}{\rm arg \ max} \ \mathcal{L}(\mathbf{z}|\mathcal{D})$.

There are a variety of acqusition functions for BO, e.g., Thompson sampling (TS), upper confidence bounds, probability of improvement and
expected improvement (EI).

The acquisition function we employ is EI that performs well and is easy to use. Suppose that we have observed $m$ points in the initialization phase of Algorithm \ref{BO} and are ready to further maximize the objective function to optimize the beam pair.
 Particularly, we wish to minimize the expected deviation from the true maximum $f(\mathbf{z}^\star)$ from previous $m$ observations, when choosing a new trial point:
\begin{equation}\label{EI}
	\begin{aligned}
	\mathbf{z}_{m+1}&= \underset{\mathbf{z}_{m+1}\in\mathcal{Z}}{\rm arg \ min} \ \mathbb{E}(\Vert f(\mathbf{z}_{m+1})-f(\mathbf{z}^\star)\Vert \ |\mathcal{D}_{m}) \\
		&\begin{aligned}=
 \underset{\mathbf{z}_{m+1}\in\mathcal{Z}}{\rm arg \ min}\int 
	& \Vert f(\mathbf{z}_{m+1})-f(\mathbf{z}^\star) \Vert \\
	& \cdot p(f(\mathbf{z}_{m+1})|\mathcal{D}_m) \  {\rm d}f(\mathbf{z}_{m+1}).
\end{aligned}
\end{aligned}
\end{equation}

However, the implement of (\ref{EI}) is expensive. In \cite{EI}, the alternative of maximizing the expected improvement in line with $f(\mathbf{z}^\star)$ was suggested. For concreteness, the improvement function is defined as
\begin{equation}
	\mathcal{I}(\mathbf{z})={\rm max} \{0,f(\mathbf{z}_{m+1})-f(\mathbf{z}^\star)\}.
\end{equation}

That is to say, when the predicted value is higher than the best value known so far, $\mathcal{I}(\mathbf{z})$ is positive;  otherwise, $\mathcal{I}(\mathbf{z})$ is zero. Therefore, the next query point can be sought by maximizing the expected improvement:
\begin{equation}
	\mathbf{z}_{m+1}=\underset{\mathbf{z}_{m+1}\in\mathcal{Z}}{\rm arg \ max} \ \mathbb{E}({\rm max} \{0,f(\mathbf{z}_{m+1})-f(\mathbf{z}^\star)\}|\mathcal{D}_m).
\end{equation}

From the normal probability density function (PDF), we can obtain the improvement likelihood of $\mathcal{I}$ relative to the normal posterior distribution characterized by $\mu (\mathbf {z})$ and $\sigma^2 (\mathbf{
z})$:
\begin{equation}
	\frac{1}{\sqrt{2\pi}\sigma(\mathbf{z})}\exp\left( 
	-\frac{(\mu(\mathbf{z})-f(\mathbf{z}^\star)-\mathcal{I})^2}{2\sigma^2(\mathbf{z})}\right) .
\end{equation}

After that, the expected improvement is evaluated by analyzing the above derivation, yielding

\begin{equation} 
	\begin{aligned}
	\mathbb{E}(\mathcal{I})&=\int_{\mathcal{I}=0}^{\mathcal{I}=\infty}	\frac{\mathcal{I}}{\sqrt{2\pi}\sigma(\mathbf{z})}\exp\left( 
	-\frac{(\mu(\mathbf{z})-f(\mathbf{z}^\star)-\mathcal{I})^2}{2\sigma^2(\mathbf{z})}\right)  
{\rm d}\mathcal{I} \\
	&=\left(\mu(\mathbf{z})-f(\mathbf{z}^\star)\right)\Phi\left(\frac{\mu(\mathbf{z})-f(\mathbf{z}^\star)}{\sigma(\mathbf{z})}\right) \\ &+\sigma(\mathbf{z})\phi\left(\frac{\mu(\mathbf{z})-f(\mathbf{z}^\star)}{\sigma(\mathbf{z})}
	\right) ,
	\end{aligned}
\end{equation}
where $\phi(\cdot)$ and $\Phi(\cdot)$ denote the PDF and cumulative distribution function (CDF) of the standard normal distribution respectively.

Finally, the EI acquisition function can be summarized as
\begin{equation}
	{\mathcal{L}}(\mathbf{z})=\left\{
	\begin{aligned}
 & 	\left(\mu(\mathbf{z})-f(\mathbf{z}^\star)\right)\Phi(Z)  +\sigma(\mathbf{z})\phi(Z), \ &\sigma(\mathbf{z})>0 \\
	 &0,  &\sigma(\mathbf{z})=0
	\end{aligned},
	\right.
\end{equation}
where $Z=\frac{\mu(\mathbf{z})-f(\mathbf{z}^\star)}{\sigma(\mathbf{z})}$.

Thus, the new sample point can be determined by maximizing $\mathcal{L}(\mathbf{z})$, as shown in the line 3 of Algorithm \ref{BA}
\section{Simulation Results}
 In this section, we carry out numerical experiments to evaluate the performance of the proposed BO based BA scheme.
 In this regard, we first reveal the correlation between the number of iterations and the spectral efficiency performance of the proposed scheme, and then compare the spectral efficiency perfomance of the orthogonal match pursuit (OMP) method, the TS-based MAB approach with BO schemes based on GP, SMAC and GBRT respectively. 
 For a good comparison, we define the normalized spectral efficiency as
 \begin{equation}
 \frac{\mathbf{SP}(X)}{\mathbf{SP}(ES)},
 \end{equation}
 where $\mathbf{SP}(X)$ presents the spectral efficiency of approach $X$ and $\mathbf{SP}(ES)$ is the spectral efficiency of exhaustive search\footnote{Exhaustive search means total sweeping in the TX and RX codebooks, which serves as the upper bound based on the codebook.}.
\begin{figure}[htbp]
	\centering
	\includegraphics[width=7.6cm,height=5.68cm]{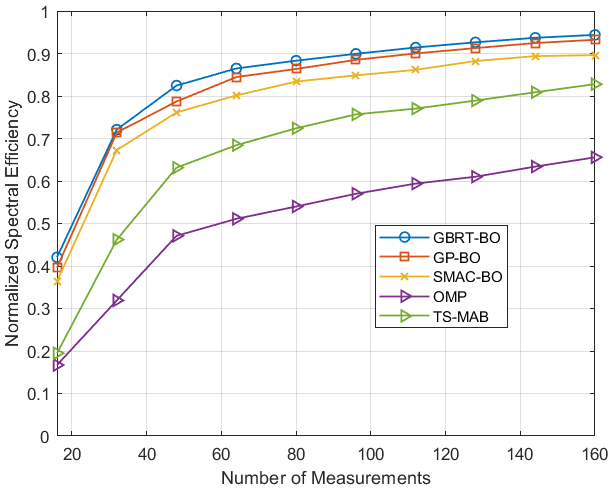}
	\centering
	\caption{Spectral efficiency of proposed schemes versus the number of measurements in the case of SNR=0 dB.   }
\end{figure}

 We consider a mmWave communication system with $N_t=64$, $N_r=16$ and $d=\lambda/2$. The discrete Fourier transform (DFT) codebook is adopted for beam codebook at the TX and the RX, and the total beam pair set size is $64\times16$, i.e., the size of codebooks for exhaustive search, OMP and TS-based MAB is set as $G_t=64$ and $G_r=16$, where each beam pair can be regarded as an arm for MAB-based methods\footnote{This implies that the performance and training iterations of MAB-based approaches are dependent on the size of codebooks. It is difficult to determine the ideal number of arms.}. This also indicates that the sensing matrix used in the OMP algorithm is the partial DFT matrix.
 For the channel setting, the propagation paths $L=5$, AoDs and AoAs are uniformly in [$-\pi/2$, $\pi/2$]. In addition, the path gains are assumed Gaussian, i.e., $\alpha_l \sim \mathcal{CN}(0,\sigma_a^2)$. In addition, the signal-to-noise ratio (SNR) is defined by SNR $=\frac{P_t\sigma_a^2}{\sigma_n^2}$.

\figurename{2} exhibits the spectral efficiency performance of OMP, MAB and the proposed schemes (GBRT-BO, GP-BO and SMAC-BO) for the cases when SNR = 0 dB and the number of measurements ranges from 16 to 160, where initial points $m=16$ and $n$ is from 1 to 144 for our proposed schemes.
In general, the selection of surrogate model is of paramount significance for BO. The simulation results have shown that GBRT based BO is more suitable for the BA problem and that BO methods outperform the OMP and MAB algorithms. Furthermore, in comparison to the non-iterative approach of CS-based methods, the iterative sweeping beam of ML-based methods has the benefit of being able to terminate the scanning beam when the performance of beam alignment achieves the criterion of initial access, hence avoiding excessive time slot waste.

\begin{figure}[htbp]
	\centering
	\includegraphics[width=7.6cm,height=5.68cm]{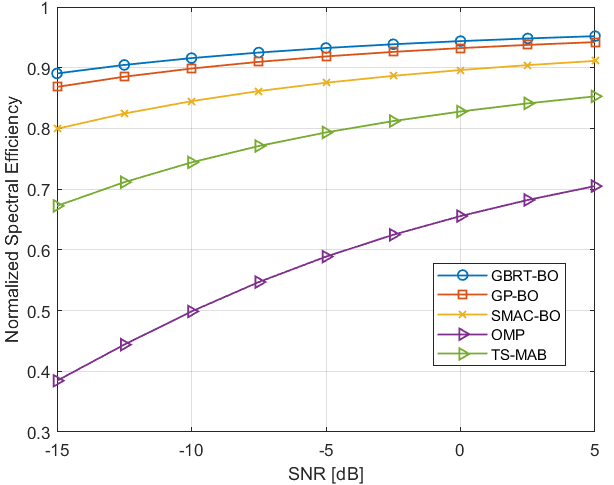}
	\centering
	\caption{Spectral efficiency of proposed schemes versus SNR in the case of the number of measurements is set to 160.  }
\end{figure}

Finally, we further investigate the impact of SNR on the spectral efficiency performance.
 \figurename{3} illustrates the spectral efficiency performance of TS-MAB, OMP and BO-based schemes in the scenario where the number of measurements is set to 160 and SNR ranges from -15 dB to 5 dB. Similar to the last experiment, BO schemes have a better spectral efficiency performance than TS-MAB and OMP methods, especially at low SNR.
\section{Conclusions}
In this study, we have presented a novel ML strategy effectively coping with BA, which regards BA as a black box problem, followed by adopting an iterative optimization method with low complexity, i.e., BO, to explore the best AoD and AoA. In addition, we suggest a GBRT surrogate model based BO and compare it with the commonly used GP and SMAC based BO schemes on the BA problem. Finally, simulation results show that BO-based schemes outperform OMP and MAB methods in terms of spectral efficiency. In our future work, we will focus on multi-objective Bayesian optimization for the multi-BS beam alignment scenario, which is a challenge for users to access the best BS.

%
\bibliographystyle{IEEEtran}
\bibliography{reference.bib}

\vspace{12pt}

\end{document}